\documentclass[superscriptaddress,twocolumn,prl,floatfix,nobibnotes]{revtex4-1}
\usepackage{graphicx}
\usepackage{amsmath}
\usepackage{bm}
\usepackage{color}
\usepackage{hyperref}

\setcitestyle{numbers,square}

\begin{document}

\title{Quantum and classical ripples in graphene}

\author{Juraj Ha\v{s}\'{i}k}
\email{jhasik@sissa.it}
\affiliation{International School for Advanced Studies (SISSA), Via Bonomea 265, I-34136 Trieste, Italy}
\affiliation{Department of Experimental Physics, Comenius University, 
Mlynsk\'{a} Dolina F2, 842 48 Bratislava, Slovakia}

\author{Erio Tosatti}
\affiliation{International School for Advanced Studies (SISSA), Via Bonomea 265, I-34136 Trieste, Italy}
\affiliation{CNR-IOM Democritos, Via Bonomea 265, I-34136 Trieste, Italy}
\affiliation{The Abdus Salam International Centre for Theoretical Physics
  (ICTP), Strada Costiera 11, I-34151 Trieste, Italy}

\author{Roman Marto\v{n}\'{a}k}
\affiliation{Department of Experimental Physics, Comenius University, 
Mlynsk\'{a} Dolina F2, 842 48 Bratislava, Slovakia}

\date{\today}

\begin{abstract}
Thermal ripples of graphene are well understood at room temperature, but their quantum counterparts at low temperatures are 
in need of a realistic quantitative description. Here we present atomistic path-integral Monte Carlo simulations of freestanding graphene, which show upon cooling a striking classical-quantum evolution of height and angular fluctuations. The crossover takes place at ever-decreasing temperatures for ever-increasing wavelengths so that a completely quantum regime is never attained. Zero-temperature quantum graphene is flatter and smoother than classical graphene at large scales, yet rougher at short scales. The angular fluctuation distribution of the normals can be quantitatively described by coexistence of two Gaussians, one classical  strongly T-dependent and one quantum about $2^{\circ}$ wide, of zero-point character. The quantum evolution of ripple-induced height and angular spread should be observable in electron diffraction in graphene and other two-dimensional materials, such as MoS$_2$, bilayer graphene, boron nitride, etc.
\end{abstract}

\maketitle

Suspended graphene is a unique physical realization of a polymerized two-dimensional (2D) membrane embedded in three-dimensional space. Its equilibrium behavior, characterized by the large distance limit $\Lambda $ of angular correlations $\langle \vec{n}(\vec{0}) \cdot \vec{n}(\vec{x}) \rangle \to \Lambda$ of tangent plane normal vectors $\vec{n}(\vec{x})$, is conventionally termed crumpled if $\Lambda$ vanishes, flat if not \cite{StatMechMembranes,Le_Doussal_Radzihovsky_Review2017}.
Like other membranes, graphene is endowed with ultra-soft out-of-plane ``flexural'' acoustic modes of dispersion $\omega=\sqrt{\frac{\kappa}{\rho}} q^2$ , where $\kappa$ is the bending rigidity, $\rho$ is the area mass density, and $q$ is the wave vector.
In the classical harmonic approximation, as a consequence of equipartition, each such mode acquires an average energy $k_B T$, leading to a height-height correlation function decaying in the long-wavelength limit as $\langle |h_q|^2 \rangle \sim q^{-4}$.
That would lead to growth of height mean-square fluctuations 
$\langle [h(\vec{x})- h(0)]^2 \rangle^{1/2} \sim |\vec{x}|$ destabilizing 
the flat phase
at any finite temperature. In reality that does not happen  because of
the long-recognized anharmonic coupling of the flexural modes to the
more regular stiffer in-plane acoustic modes 
\cite{Nelson_Peliti1987}, a coupling reflecting the physical necessity
for the membrane's planar extension to shrink when corrugated.  Its
effect on the classical thermal equilibrium state of the membrane was
studied by renormalization-group (RG) methods \cite{Nelson_Peliti1987,
  PhysRevB.89.125428, PhysRevB.89.224307,
  Le_Doussal_Radzihovsky_Review2017} and by classical Monte Carlo
(MC) \cite{Fasolino2007, PhysRevLett.116.015901, Phys.Rev.B.80.121405}
and molecular dynamics (MD) simulations \cite{doi:10.1063/1.4897255}
leading, despite some variants \cite{PhysRevB.93.235419}, to a
consistent picture. Anharmonic coupling increases the $ q \rightarrow
0$ bending rigidity from the constant $\kappa$ to $q$-dependent
$\kappa(q) \sim q^{-\eta}$ which stiffens the membrane and
renormalizes the long-wavelength height fluctuations $\langle |h_q|^2
\rangle$ from $\langle |h_q|^2 \rangle \sim q^{-4}$ to $\langle
|h_q|^2 \rangle \sim q^{-4+{\eta}}$.  With $\eta \sim 0.8-0.85 $,
the state of the membrane is restored from crumpled to flat
\cite{Le_Doussal_Radzihovsky_Review2017}.

So far we discussed the known classical picture. The low-temperature quantum
membrane must behave differently. Already at the harmonic level the
$\omega$-independent equipartition energy $k_B T$ per mode is replaced
by the zero-point energy $\frac{1}{2} \hbar \omega$. That yields the
very different height-height correlation function $\langle|h_q|^2
\rangle \sim q^{-2}$ (Ref.\cite{PhysRevB.89.125428}), representing a
flat quantum ground state as opposed to the crumpled classical
harmonic state at finite temperature~\cite{PhysRevE.94.032125}.
Compared to thermal fluctuations, quantum zero-point motion implies
weaker long-wavelength flexural fluctuations, yet relatively enhanced
short-wavelength ones, suggesting that quantum ripples will make
graphene globally flatter but locally rougher than their classical
counterpart. The actual rippling of freestanding graphene including quantum fluctuations and anharmonic interaction between the modes was
addressed in a series of recent theoretical approaches 
\cite{PhysRevB.89.224307,PhysRevE.94.032125,PhysRevB.89.125433,PhysRevB.94.079904}, whose predictions are limited to the long-wavelength behaviour of
correlations. A complementary study of quantum and anharmonic effects on the flexular mode was performed in Ref. \cite{PhysRevB.91.134302} by lattice dynamical methods. Path-integral Monte Carlo (PIMC) \cite{PhysRevB.92.195416} and path-integral molecular dynamics simulations \cite{Herrero-Ramirez2016, Herrero-Ramirez2017, Herrero-Ramirez2018} addressed at the atomistic level some important thermodynamic quantities. We address here the quantities of crucial physical interest, the height-height correlation function and the angular distribution of surface normals.

Here we present extensive PIMC atomistic
simulations, whose results, besides agreeing with RG asymptotics, provide also a fully realistic description of height correlations and angular profiles of graphene under quantum and thermal fluctuations, whose evolution from classical to quantum is revealing in view of ongoing experimental tests.

Graphene was modeled by a fully mobile honeycomb lattice of $N$
pointlike atoms (initially $N=19\ 440$, scaling down to 4860 
for quantum PIMC at very low temperatures, and up to 108 864 for classical MC) interacting through Tersoff's
empirical potential \cite{PhysRevB.37.6991,PhysRevB.81.205441}. This potential was optimized to reproduce experimental phonon-dispersion relations of in-plane graphite and is computationally less demanding than the reactive bond-order potential \cite{Phys.Rev.B.72.214102} used in the series of MC simulations \cite{Fasolino2007, PhysRevLett.116.015901, Phys.Rev.B.80.121405}.
The single 2D membrane is replaced in PIMC by imaginary-time-coupled copies
(Trotter slices), whose number $M$ must be large enough at each given
temperature $T$ to warrant convergence. Details about simulations, the PIMC method \cite{PhysRevB.51.2723, PhysRevE.57.2425, RevModPhys.67.279, Herrero-Ramirez2014}, and Trotter convergence are further discussed in the Supplemental Material (SM) \cite{suppmat}.

\begin{figure}[tp]
  \includegraphics*[width=\columnwidth]{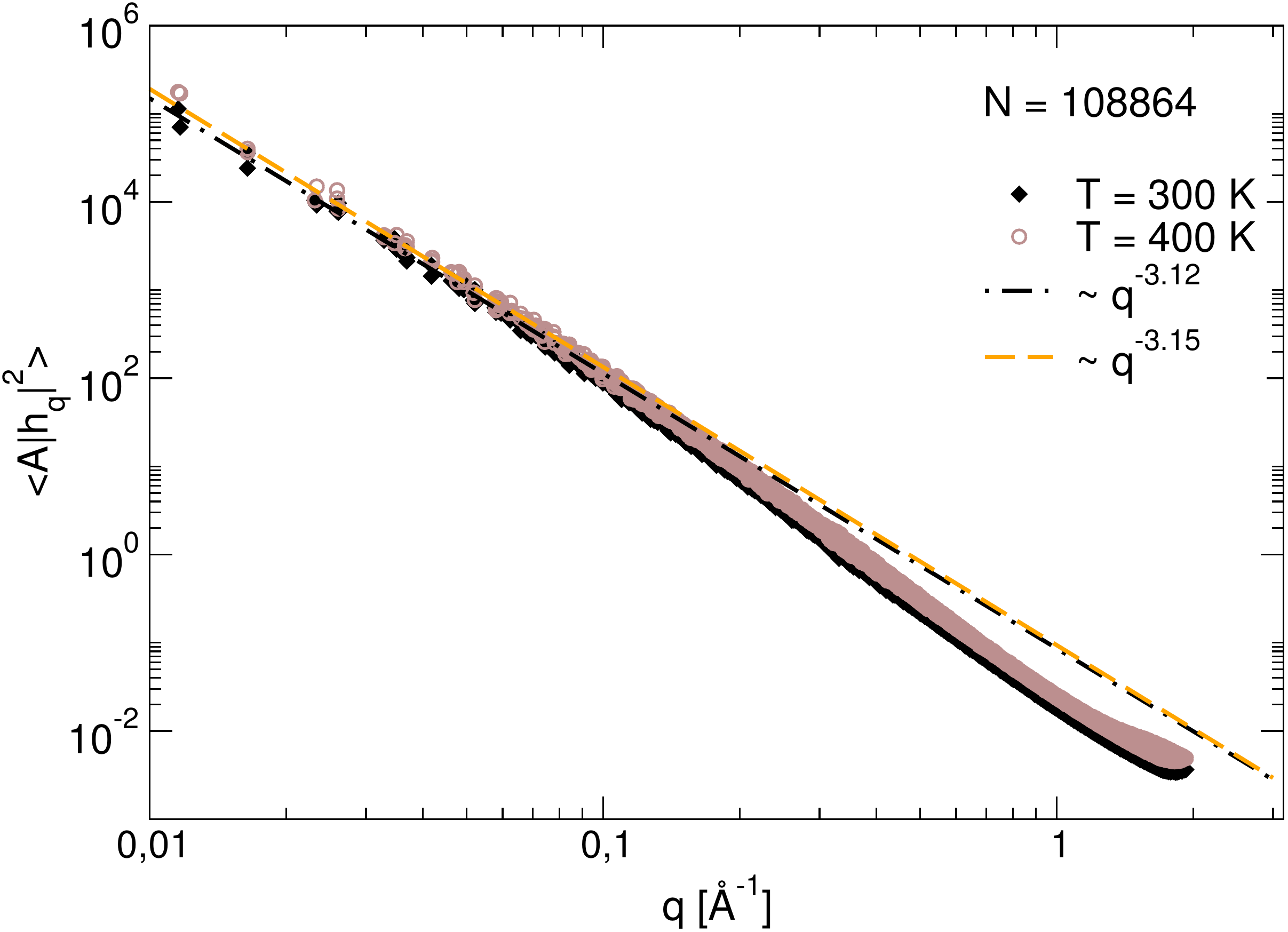}
  \caption{Classical graphene height-height correlation function for
    $N=108\ 864$. The long-wavelength asymptotic slope $-4+\eta$ equals
    $-3.12 \pm 0.05$ at 300 K and $-3.15 \pm 0.05$ at 400 K, 
    $\eta$ reflecting the effect of anharmonic fluctuations. This can be compared 	  to the classical behaviour for the smaller system $N=4860$ at $T=0.6$ K shown     	in Fig. \ref{fig:PIMC_0.6K} where the slope is $-3.98$, as expected for purely
    harmonic fluctuations.}
 \label{fig:MC}
\end{figure}

As an initial test of the MC simulation method we first calculated the
height-height correlation function in the classical case at $T=300$
and 400 K. The results for $N=108\ 864$ (linear size 530 \AA) are shown in
Fig. \ref{fig:MC} (in this and all subsequent figures the height-height correlation function is normalized by the system area $A$, see Computational Methods, SM).  The long-wavelength height fluctuations show two distinct regimes, crossing over at $q \sim 0.1$ \AA$^{-1}$ from harmonic
$q^{-4}$ to anharmonic $q^{-4+\eta}$ with $\eta = 0.85 - 0.88$, in
good agreement with RG predictions, and with numerical
simulations~\cite{PhysRevE.94.032125, Le_Doussal_Radzihovsky_Review2017, Phys.Rev.B.80.121405}; claims that the classical correlation function should behave for $q \rightarrow 0$ as $\langle|h_q|^2 \rangle \sim q^{-2}$
~\cite{PhysRevB.93.235419} are therefore not substantiated.

\begin{figure}[tp]
  \includegraphics*[width=\columnwidth]{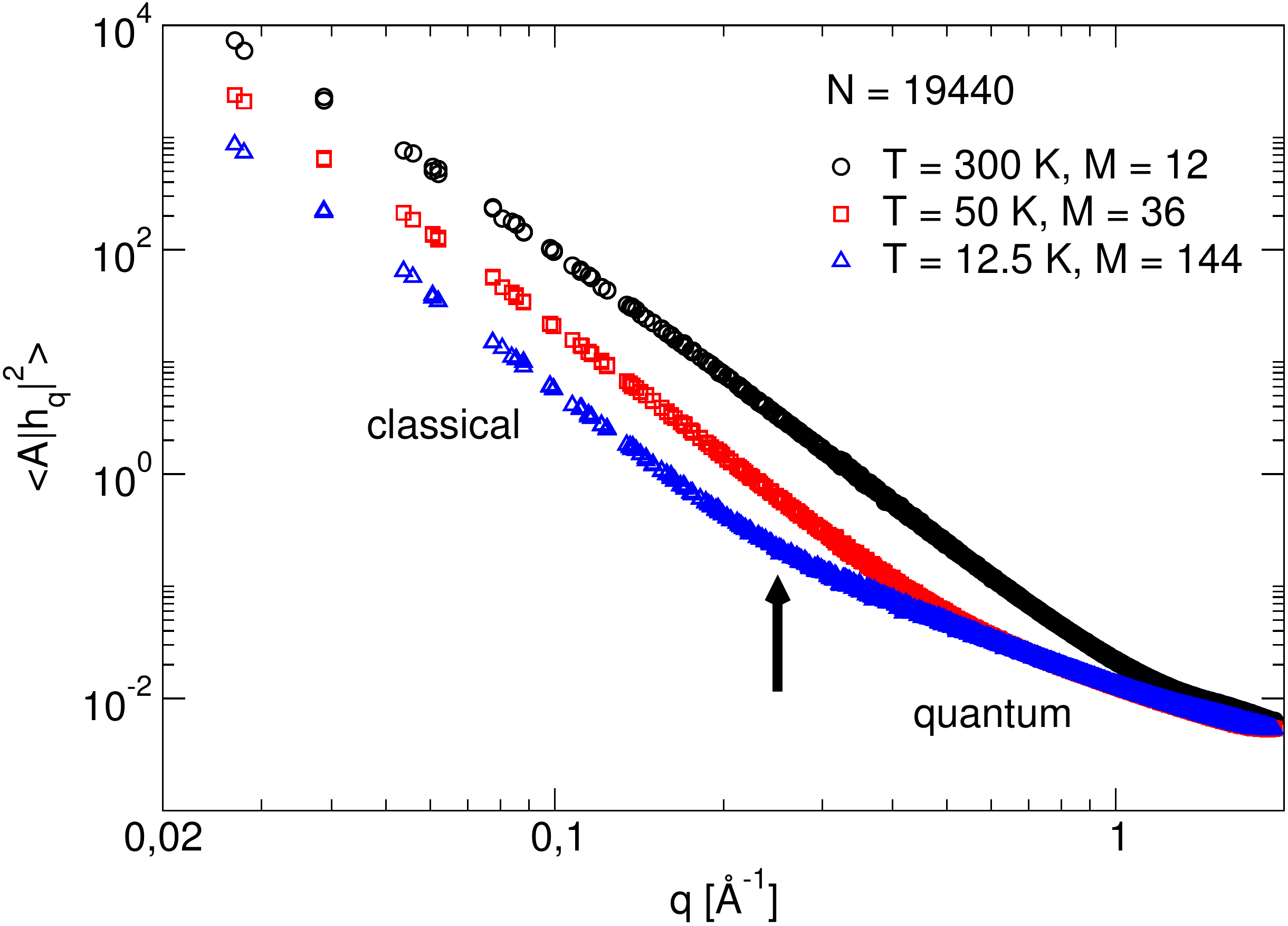}
  \caption{Graphene PIMC height correlations for $N=19\ 440$. The arrow marks the 	  crossover from thermal to quantum regime for $T = 12.5$ K. The classical
    regime is visible at small $q$, whereas the quantum regime emerges at large 	$q$.}
 \label{fig:N19440PIMCgzz}
\end{figure}

\begin{figure}[hbp]
  \includegraphics*[width=\columnwidth]{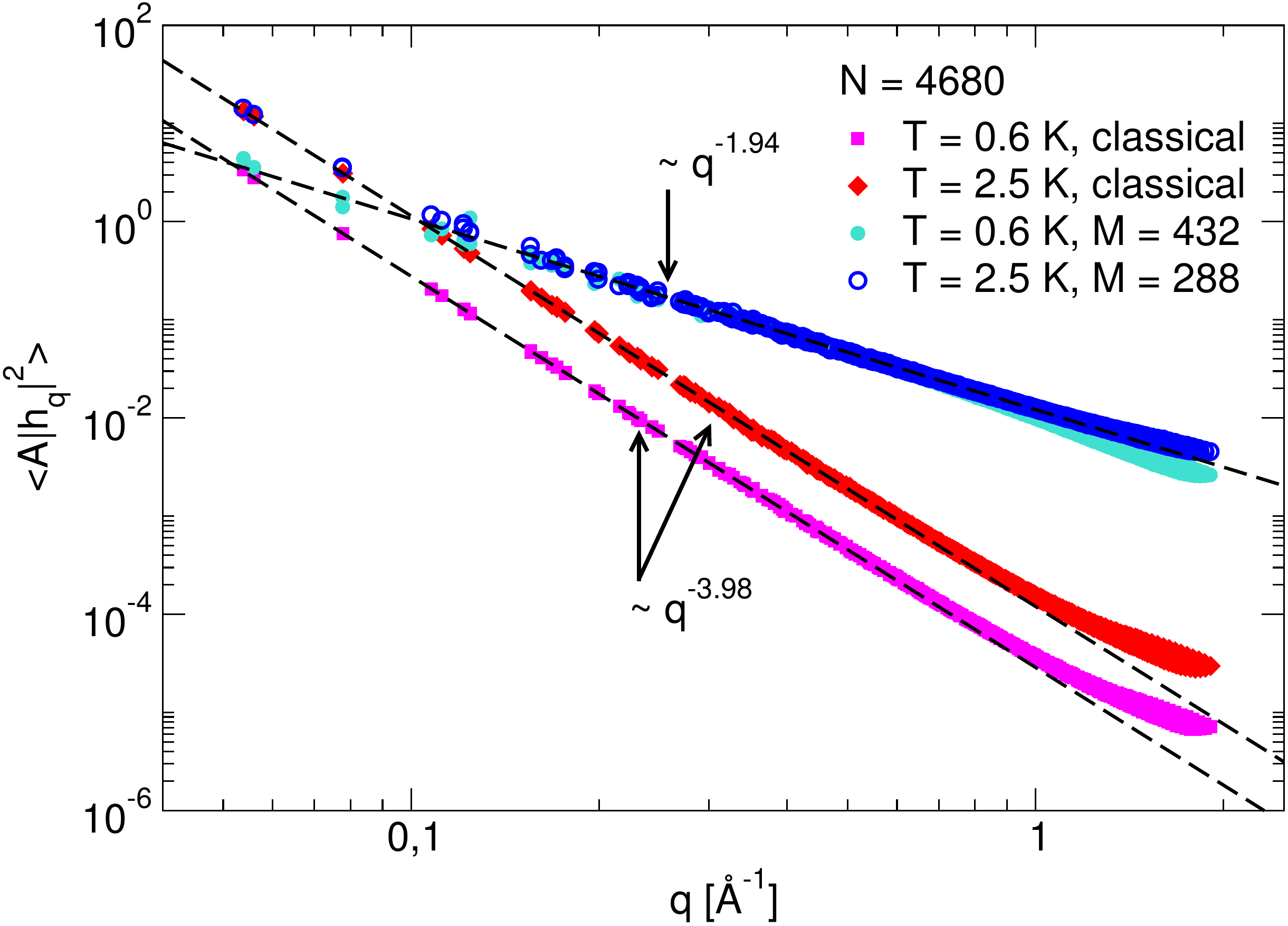}
  \caption{Height-height correlations for $N=4860$ 
  at $T=0.6$ and $T=2.5$ K: classical and PIMC ($M=288$ and $M=432$). The
    region $q < 0.6$ \AA$^{-1}$ is fully converged with respect to the Trotter
    number $M$. The slope of the classical curves is $-3.98 \pm 0.003$ while that
    of the quantum curves is $-1.94 \pm 0.02$ for $q > q_T$.}
 \label{fig:PIMC_0.6K}
\end{figure}

Satisfied by this check we moved on to main objective of our 
work
- quantum PIMC simulations. In anticipation of increasing system size along the Trotter dimension, we start with a smaller membrane of $N=19\ 440$ atoms and proceed by cooling it down. Figure \ref{fig:N19440PIMCgzz} shows the PIMC height correlation function for $N=19\ 440$ (220 \AA~linear size) at $T=$ 300, 50 and 12.5 K. Both $T=50$ and 12.5 K data show a clear crossover from classical
correlations for $q < q_T$ to quantum ones taking place, again in
agreement with predictions, for $q > q_T$, where
~\cite{PhysicsReports2015, PhysRevE.94.032125}
\begin{equation} 
q_T =  \left(\sqrt{\frac{\rho}{\kappa}} \frac{2 k_{B} T}{\hbar}\right)^{1/2} 
\label{qT}
\end{equation}
marks the point where the classical equipartition energy of the
flexural phonon equals its quantum zero-point energy, $\frac{1}{2}
\hbar \omega \approx k_B T$.  Curves at $T=50$ and 12.5 K collapse
for $q > 0.5$ \AA$^{-1}$ but diverge for lower $q$-vectors. The quantum
result 
representing  
the $T=0$ correlations is gradually emerging as the
lower envelope of finite-temperature curves at large $q$. For this
relatively large system, however, it is hard to push temperature down
any further. 
At $T=12.5$ K the total number of simulated particles is $144 \times 19\ 440 \approx 3\times 10^6$.
%

\begin{figure}[tp]
  \includegraphics*[width=\columnwidth]{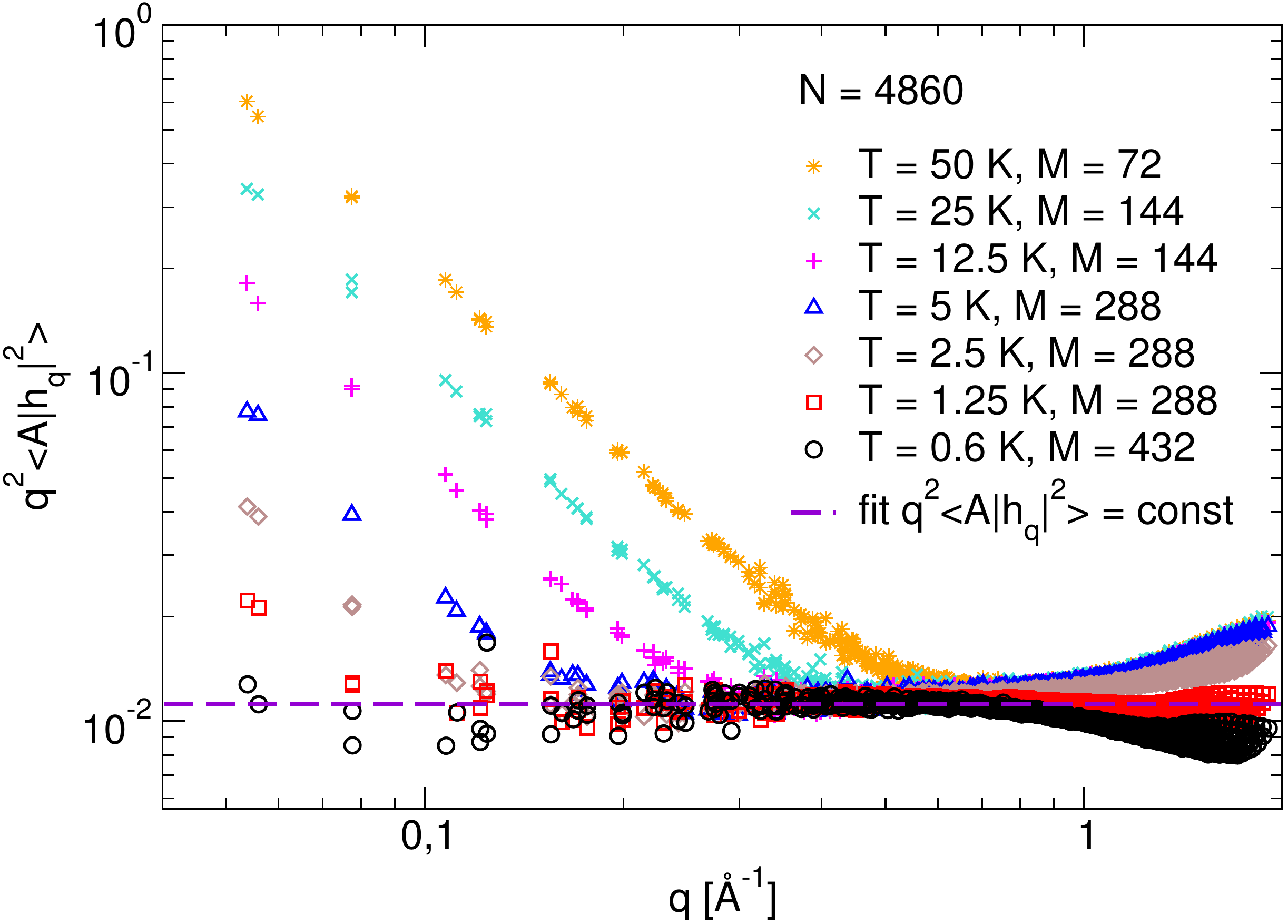}
  \caption{Temperature-dependent $N=4860$ PIMC height correlations
    of graphene, multiplied by $q^2$ so that $q^{-2}$ behaviour
    appears as horizontal. The gradual temperature-dependent veering
    from quantum correlations towards classical ones at longer
    wavelengths is clearly visible. The violet horizontal dashed line represents the best fit of $T = 0.6$ K data in the region $q < 0.6$ \AA\ to the quantum harmonic approximation prediction $q^2 A \langle|h_q|^2\rangle =$ const and serves as a guide to the eye.}
 \label{fig:PIMC}
\end{figure}

By halving the linear size to $L=110$ \AA\ with $N=4860$ (lowest-$q$-value 
$q_{min} = 2\pi/L \sim 0.057$ \AA$^{-1}$), we could reach
much lower temperatures down to $T=0.6$ K. The resulting classical and
PIMC height-height correlations at $T=2.5$ and $T=0.6$ K are
shown in Fig. \ref{fig:PIMC_0.6K}, the first striking result of this
paper. 
It can be seen that the PIMC curves at both temperatures coincide
(apart from region $q > 0.6$ \AA$^{-1}$ where the Trotter convergence is
not perfect) down to $q \sim 0.1$ \AA$^{-1}$ where at $T=2.5$ K the
classical behaviour takes over. On the other hand, the PIMC curve at
$T=0.6$ K shows the quantum regime in the whole region of $q$-vectors
spanning about one decade. The PIMC and classical curves are very
different, the quantum slope very close to the harmonic
value of 2, compared to about 4 of the classical case.  The PIMC
height correlations for all simulated temperatures (now multiplied by
$q^2$ for convenience) are shown in Fig. \ref{fig:PIMC}. Ignoring
points above $q \approx 0.6$ \AA$^{-1}$, whose scatter is caused by
incomplete Trotter convergence, two regimes are again visible, large
$q$ and small $q$. The first shows a flat horizontal behaviour,
similar to that expected in the quantum harmonic case, $q^{2}
\langle|h_q|^2 \rangle \sim$ const. The second, for low $q$,
represents the classical regime. The crossover between the two is
again at $q=q_T$. At the lowest temperature of 0.6 K the quantum
behaviour therefore extends down to the smallest $q$-vector $q \approx 0.057$
\AA$^{-1}$.  One can conclude that at any given temperature $T$ the
ripples of graphene are quantum and of moderate amplitude for shorter
wavelengths than $\lambda_T = 2\pi/q_T \sim T^{-(1/2)}$,
stronger and classical for longer wavelengths, but never disappearing
at any finite temperatures. The temperature dependence of the
crossover $q_T$ extracted from Fig. \ref{fig:PIMC} is shown in the
SM, Fig. 4. Our results suggest, in agreement with
Ref. \cite{PhysRevE.94.032125} that quantum fluctuations leave the
membrane flat on the large scale. The interesting logarithmic corrections
implied by D = 2 being the upper critical dimension of quantum
membranes \cite{PhysRevE.94.032125} cannot be addressed here because of
computational limitations.  Accessing even lower $q$ requires both
larger system size and lower temperatures, which in turn imply higher
Trotter number, and such simulations appear at the moment
prohibitively expensive.

\begin{figure}[tp]
\includegraphics*[width=\columnwidth]{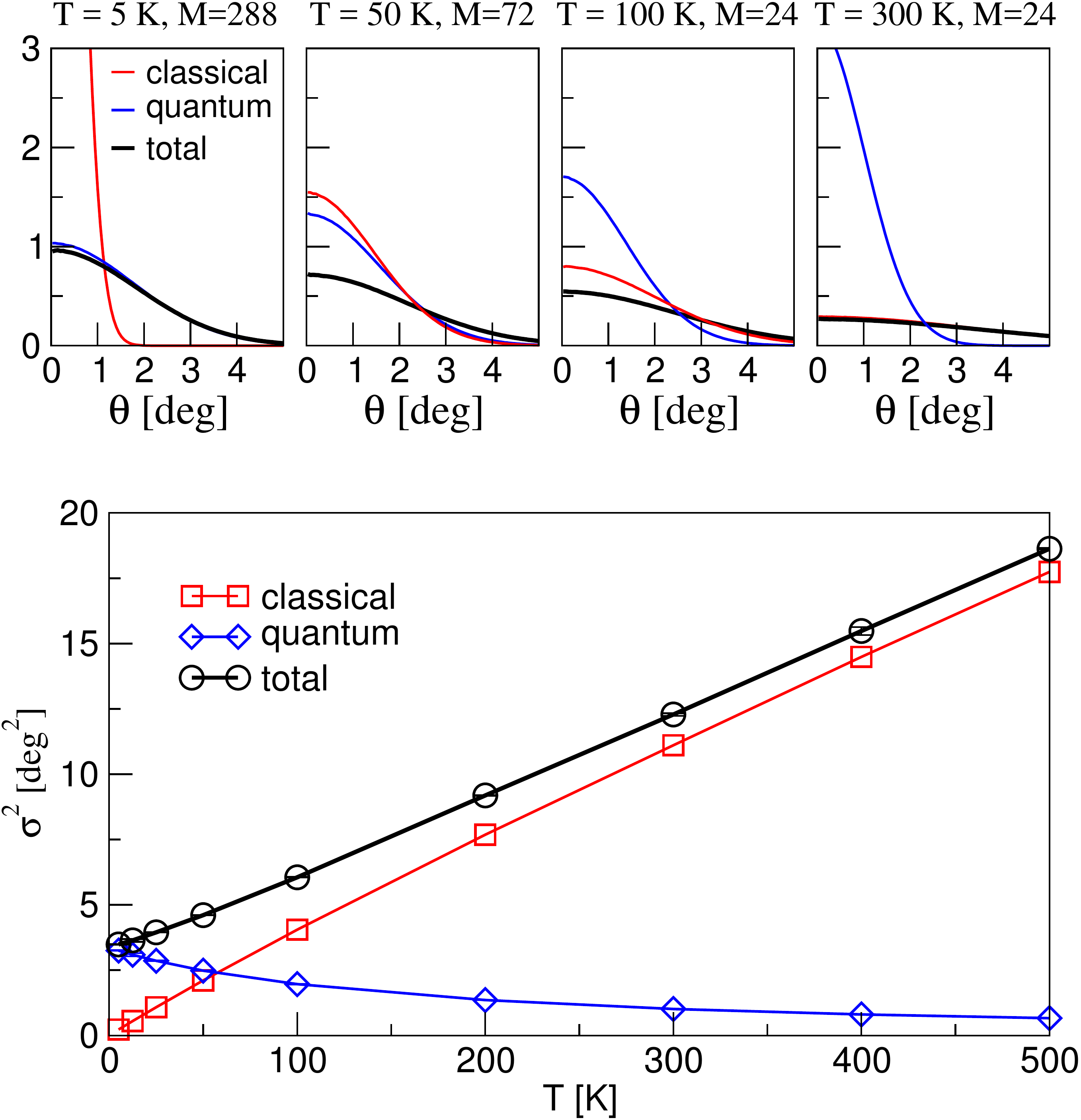}
\caption{(Top) Angular probability distribution of graphene's normal for
  $N = 4860$, decomposed into classical and quantum contribution for
  increasing temperatures. (Bottom) Temperature dependence of the total
  variance $\sigma^2$ of the distribution of normal angles and its classical
  and quantum components. The variance was determined by fitting the
  curves to the Gaussian form $C \exp(-\frac{\theta^2}{2 \sigma^2})$.}
\label{fig:normals}
\end{figure}

A complementary feature and consequence of graphene's ripples, accessible to measurement together with height correlations, is provided by the angular
distribution $P(\theta)$ of $\theta$, defined for each atom as the
angle between the normal to the triangle formed by its three
neighbours and the normal to the averaged membrane plane.  On this
atomistic scale, all fluctuations including those at large $q$
simultaneously contribute to the angular spread. Classically,
$P(\theta)$ should evolve from $\delta(\theta)$ at $T = 0$ to a
Gaussian of finite width at $T>0$. Whereas PIMC simulations fully
account for both classical and quantum contributions to the angular
spread, they also allow very naturally to decompose the total averaged 
fluctuations in the two distinct contributions. Generally, the
classical contribution is given by the fluctuations of the centroids of Trotter beads averaged over all imaginary time slices whereas the quantum contribution is represented by the fluctuations around the centroids within the slice. In our case we calculated first the classical normal angle
distribution for the configuration given by centroids; the angle
between the normal in each Trotter slice and the classical one
represents the quantum contribution. The analysis of the probability distribution
$P(\theta)$ represents the second important result of this 
paper. 
The distributions of total, classical, and quantum normal angle fluctuations 
calculated at several temperatures from 5 to 300 K are shown in 
Fig. \ref{fig:normals} (upper panel).  All of them are rather accurately 
Gaussian, with a $T$-dependent width or spread $\sigma$. At $T=5$ K 
the total spread is dominated by the quantum contribution; at 50 K classical and quantum compete; and at 300 K the classical prevails over the (declining) quantum.

\begin{figure}[tp]
\includegraphics*[width=\columnwidth]{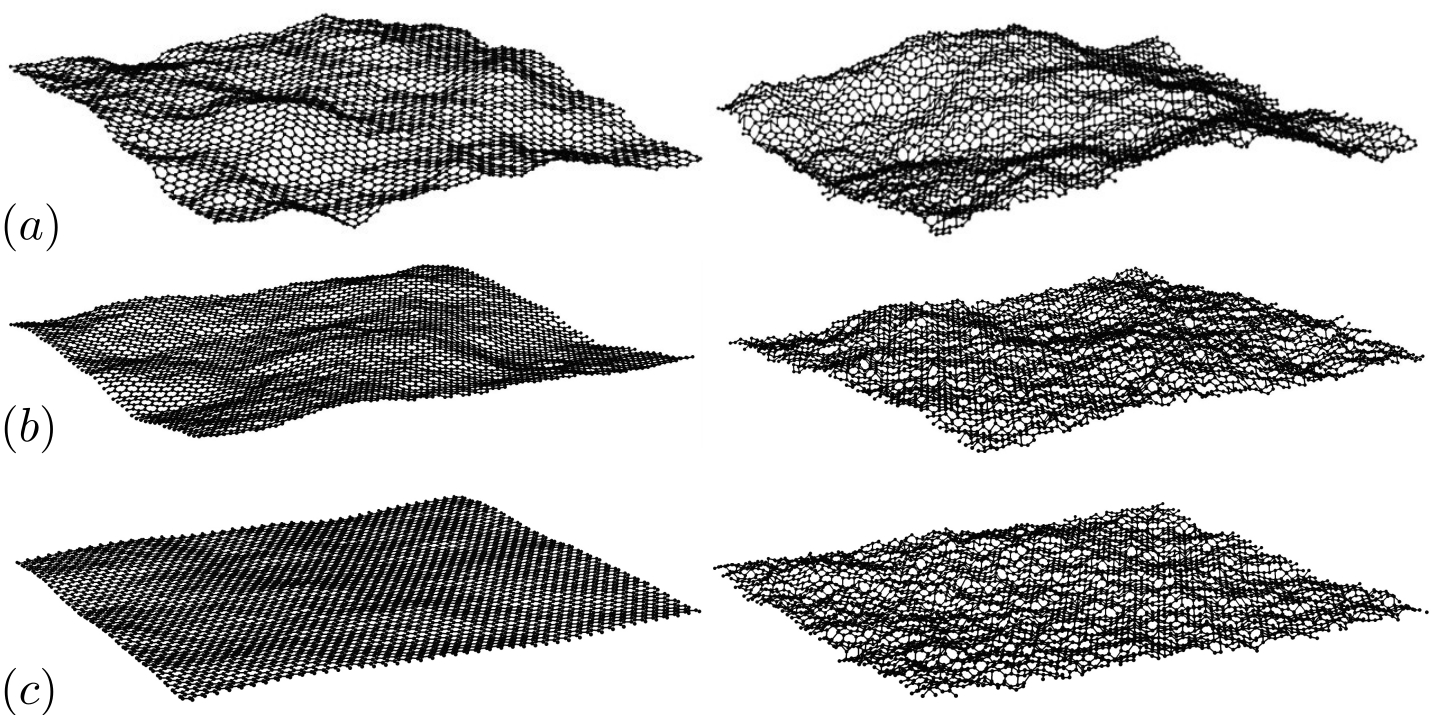}\\
\caption{Comparison of configurations from classical simulations (left) with
  a single Trotter slice from the PIMC with M Trotter slices (right) for $N = 4860$ at
  different temperatures. (a) $T = 50$ K, $M = 72$; (b) $T = 12.5$ K,
  $M = 144$; (c) $T = 2.5$ K, $M=288$. For convenience, out-of-plane
  displacements are amplified by a factor 10.}
\label{fig:CLSvsPIMCconf}
\end{figure}

The crossover between classical and quantum behaviour for normal angle fluctuations is thus around 50 K, slightly lower than expected for some other                                                                                                                                                                                                                                                                                                                                                                                                                                                                                                                                                         
thermodynamic quantities ~\cite{PhysRevB.89.224307}. 
Figure \ref{fig:normals} (lower panel) shows
the temperature dependence of the spread of all distributions. Whereas at
room temperature the quantum spread is much smaller than the classical
one, at $T=50$ K the quantum one takes over upon cooling, saturating
at very low temperatures at a value of $\sigma$ of roughly $2^{\circ}$.
The histogram of normal angles at $T=5$ K including the $\sin \theta$
factor from the solid angle element (see the SM, Fig. 5) shows
that the normal angle fluctuations reach up to $4^{\circ}$ and the
most likely value of the angle $\theta$ is around $\sigma =
1.8^{\circ}$.  This substantial residual angular spread represents a
rather spectacular manifestation of quantum zero-point motion in a
macroscopic membrane.

A pictorial idea of the difference between classical and quantum
graphene at different temperatures is offered by simulation snapshots
for a variety of cases (Fig. \ref{fig:CLSvsPIMCconf}) of the $N=4860$
membrane.  In the bottom panel (c) we see that at $T=2.5$ K the
classical membrane would be nearly flat with just small
long-wavelength ripples. Quantum graphene, which at that temperature
is close to its ground state, sports instead sizable short-wavelength
ripples, which make it locally much rougher.  This roughness is also
responsible for the broader angular distribution of the normals shown
in Fig. \ref{fig:normals}. In the middle [(b), $T=12.5$ K] and top
[(a), $T=50$ K] panels, classical graphene is seen to progressively
develop long-wavelength ripples and so does the quantum one on top of
the local roughness.

The $q$-dependent character of the classical/quantum crossover in
graphene requires low temperatures for longer length scales to an
extent dictated by its low bending rigidity $\kappa \sim 1 - 2$ eV
(the precise value is still debated \cite{PhysRevB.91.134302}).
In this respect, other 2D monolayers should display their
quantum-classical crossover, albeit quantitatively less dramatic, at
higher temperatures, owing to their larger values of $\kappa$. For
example, the MoS$_2$ monolayer has an area density four times higher
but a bending rigidity $\kappa \sim 9 - 10$ eV~\cite{MoS2-rigidity} 
and might therefore be more suitable than graphene for observation of
quantum ripples at high temperature, due to larger value of the
coefficient $\sqrt{\frac{\kappa}{\rho}}$ in the flexural mode
dispersion law [see Eq.(\ref{qT})]. The graphene bilayer could also be
of interest in this respect -- bending rigidity data in literature,
however, scatter considerably, from 3.35 \cite{doi:10.1063/1.4915075} to 35 eV \cite{Lindahl2012}.

The atomistic height correlations and angular fluctuations as well as the thermal
crossover just predicted are directly amenable to experimental
verification, particularly through elastic scattering of impinging
particles or radiation. Surface normal fluctuations of several degrees
were observed in freely suspended graphene sheets at room temperature
by transmission electron microscopy \cite{Meyer2007}. For micron-size
areas such as those realizable for suspended graphene, electron
scattering by micro-low-energy electron diffraction/low-energy electron microscopy is already available -- so far only at temperatures of 150 K and higher, that is, above the quantum-classical crossover~\cite{PhysRevB.78.201408, Morgante2010,
  PhysRevB.84.115401}. The intrinsic part of the diffraction peak
width observed in these experiments seems indeed of the order of
several degrees in monolayer graphene, and shrinks as expected on
cooling from 450 to 150 K. We predict therefore that the shrinking
should cease at a crossover temperature around 50 K to a residual
zero-point width of the order $2^\circ$.  At the same time, the roughness
spectral composition should shift from longer to shorter wavelengths
as indicated by Fig.\ref{fig:PIMC}.

To summarize, PIMC simulations provide a quantitative atomistic
description of the influence of quantum mechanics on the ripples of
freestanding graphene. Graphene or other kinds of 2D membranes offer
a unique possibility to compare and visualize quantum and classical
fluctuations in a 2D condensed-matter system. Quantum fluctuations
make graphene flatter at long wavelengths, in agreement with RG
studies\cite{PhysRevB.89.224307, PhysRevE.94.032125}. In the region of
$q$-vectors spanning one decade down to $q \approx 0.06$ \AA$^{-1}$,
corresponding to the length scale of about 100 \AA, we found the 
height-height correlation function to follow the harmonic prediction
$q^{-2}$, marginal logarithmic corrections~\cite{PhysRevE.94.032125}
being invisible on this scale. The angular distribution of normal angles
exhibits a zero-point-induced spread of several degrees predicted to
persist below a crossover temperature of about 50 K which will be
fully measurable by scattering techniques. Other 2D systems, such as
graphene bilayer or MoS$_2$ with larger bending rigidities and
therefore with quantitatively smaller spreads, will have higher
crossover temperatures. Quantum ripples should also be relevant to low-temperature graphene tribology. Finally, the flexural fluctuations and consequent local quantum roughness are likely to affect the electronic properties of graphene at low temperatures. The importance of ripples for electron-phonon coupling was demonstrated, e.g., in Ref. \cite{Laitinen2014}. Further investigations in this direction stand as an interesting problem to be considered in the future.

\begin{acknowledgments}
J. H. and E.T. would like to thank A. Morgante, S. Sorella and F. Becca for 
stimulating discussions. R.M. was supported by the VEGA project No.~1/0904/15 and by the Slovak Research and Development Agency under Contract No.~APVV-15-0496. The calculations were performed in the Computing Centre of the Slovak Academy of Sciences using the supercomputing infrastructure acquired in Projects No. ITMS 26230120002 and No. 26210120002 (Slovak infrastructure for high-performance computing) supported by the Research \& Development Operational Programme funded by the ERDF. Work in Trieste was sponsored by ERC Advanced Grant 320796 - MODPHYSFRICT.
\end{acknowledgments}

\bibliographystyle{apsrev4-1}

\end{document}